\documentclass[11pt,copyright,creativecommons,noderiv,noncommercial]{eptcs}
\usepackage[english]{babel}

\usepackage{breakurl}
\usepackage{amsmath}
\usepackage{amssymb}
\usepackage{semantic}
\usepackage{braket}
\usepackage{varwidth}

\newcommand{\Int}{\mathcal{Z}}

\newcommand{\syntax}[1]{\mathtt{#1}}
\newcommand{\links}{\textsc{Links}}
\newcommand{\tinylinks}{\textsc{TinyLinks}}

\newcommand{\checks}[1]{\stackrel{#1}{<=}}
\newcommand{\syn}[1]{\stackrel{#1}{=>}}
\newcommand{\checkval}{\checks{val}}
\newcommand{\synval}{\syn{val}}
\newcommand{\checkexp}{\checks{exp}}
\newcommand{\synexp}{\syn{exp}}

\newcommand{\power}[1]{\wp(#1)}

\newcommand{\indlist}[3]{#1_{#2},\,\ldots,\,#1_{#3}}
\newcommand{\tindlist}[4]{#1_{#3}:#2_{#3},\,\ldots,\,#1_{#4}:#2_{#4}}
\newcommand{\forallIndex}[3]{\forall #1 \in \Set{#2,\: \ldots\:, #3}}
\newcommand{\tuple}[2]{\left(#1,\, \ldots\,, #2\right)}
\newcommand{\tupleInd}[3]{\tuple{#1_{#2}}{#1_{#3}}}
\newcommand{\bind}[3]{#1\left[ #3/#2 \right]}
\newcommand{\slist}[1]{\left[\,#1\,\right]}
\newcommand{\lift}[1]{\left\lfloor\, #1\, \right\rfloor}

\newcommand{\sem}[2]{#1|[\syntax{#2}|]}
\newcommand{\vsem}[3]{\sem{\mathcal{V}}{#1}#2\,#3}

\newcommand{\avsem}[3]{\mathcal{V}|[\syntax{#1}|]^a #2\,#3}
\newcommand{\ssem}[3]{|[\syntax{#1}|]#2\,#3}

\newcommand{\assem}[3]{|[\syntax{#1}|]^a #2\,#3}

\newcommand{\eenvTotpred}{eenvToTpred}
\newcommand{\code}[1]{\texttt{#1}}

\title{An Abstract Semantics for Inference of Types and Effects in a Multi-Tier Web Language}
\author{Letterio Galletta \qquad\quad Giorgio Levi\institute{Dipartimento di Informatica, Universit\`a di Pisa} \email{\{galletta,\,levi\}di.unipi.it}}

\def\titlerunning{An abstract semantics for inference of types and effects in a multi-tier web language}
\def\authorrunning{L. Galletta, G. Levi}

\begin{document}
  
\maketitle
  
\begin{abstract}
Types-and-effects are type systems, which allow one to express general semantic properties and to statically reason about program's
execution. They have been widely exploited to specify static ana\-lyses, for example to track computational side effects, exceptions and communications
in concurrent programs. 
In this paper we adopt abstract interpretation techniques to reconstruct (following the Cousot's methodology) a types-and-effects system developed
to handle security problems of a multi-tier web language. Our reconstruction allows us to show that this types-and-effects system is not
sound with respect to the semantics of the language. In addition, we correct the soundness issues in the analysis and  
systematically construct a correct analyser.
\end{abstract}

\section{Introduction}

Types-and-effects systems are a powerful extension of type systems which allows one to express general semantic properties and to statically reason 
about program's execution. The underlying idea is to refine the type information so as to express further intensional or extensional properties of
the semantics of the program: in practice, they compute the type of each program's sentence and an approximate (but sound) description of its run-time
behavior. Since they are defined over the well understood theory of type systems, they are an intuitive framework for specifying and for developing
static analyses. 
Such systems were originally introduced in \cite{Gifford:1986:IFI:319838.319848} to statically track side effects in languages that mix functional and imperative feature. 
However, they have been employed to control many other kinds of computational effects and analyses, e.g.~exceptions \cite{Leroy:2000:TAU:349214.349230}, 
region inference \cite{NiNi05ppa} and communications in concurrent programs \cite{springerlink:10.1007/BFb0016845}. Recently, they have been used in \cite{baltopoulos09} 
to handle security issues in \links{} \cite{Cooper06links}. 

\links{} is a strict, typed, functional language for web applications. Its main feature is to be multi-tier, that is, it enables the developer to 
mix client, server and database source code by delegating the charge of code and data partitioning to the compiler: from a single source file the 
compiler generates code for the database back-end, for the web server and the client front-end, ensuring that all data is stored either in client
or in database. In \cite{baltopoulos09} Baltopoulos and Gordon have shown that storing unencrypted application data on the client opens \links{} to attacks
that may expose secrets and modify control-flow and application data. In order to overcome these problems they have proposed a compilation strategy based on
authenticated encryption\footnote{a combination of secrecy and integrity protection obtained by encrypting together data and its hash.} 
and a types-and-effects system to enforce programs to satisfy a particular class of integrity constraints (event-based assertions).
This types-and-effects system formalizes source level reasoning about \links{} programs and allows them to prove security properties by inspection of 
the source code. For the definition of this system they have followed a methodology characterized by translating each \links{} expression to an expression 
of a concurrent $\lambda$-calculus with refinement types \cite{Bengtson:2008:RTS:1380848.1381243}. This translation hides the properties of the analysis, and does not guarantee the
soundness with respect to the semantics of the language. Hence, we decided to study the properties of this analysis by reconstructing it by abstract interpretation 
\cite{galletta10}.

Abstract interpretation \cite{CousotCousot77-1,CousotCousot79-1,CousotCousot92-1,CousotCousot92-2} is a general theory for approximating the semantics of dynamic systems.
The key idea behind abstract interpretation is that the description of the behavior of a system (at various levels of abstraction)
is an approximation of its formal semantics. 
In static analysis this means that every property of a program can be observed in its semantics 
and computed as an approximation: the intuition is that the analysis can be systematically derived by throwing away superfluous information from 
the semantics. In practice, the approximated semantics (abstract semantics) is obtained 
from the standard one (called concrete) by substituting the actual (concrete) domain of computation and its basic semantic operations
with abstract domain and abstract semantic operations, respectively. 
The basic idea is that the abstract domain is a representation 
of some properties of interest about concrete domain's values, while abstract operations simulate, over the abstract domain, the behavior of 
their concrete counterparts. Hence the abstract semantics computes the properties of interest and  
the analysis algorithm corresponds to evaluating programs over the abstract domain. 
Since the abstract domain is a sound approximation of the concrete one, 
the analysis algorithm is correct with respect to the semantics by construction.

Type systems (and corresponding type inference algorithms) have been reconstructed as a hierarchy of abstract interpretations by Cousot \cite{Cousot97-1}.
In order to reconstruct the types-and-effects analysis of \links{} we extend Cousot's methodology by defining an abstract domain able to express 
types augmented by effects. In this paper we give the following contributions:
\begin{itemize}
 \item we demonstrate that the analysis defined by Baltopoulos and Gordon is not sound: in fact, the expression 
 $\mathrel{\syntax{get(}}\mathrel{\syntax{Text(}}\syntax{"Hello!"}\mathrel{\syntax{)}}\mathrel{\syntax{)}}$ is type-checked but it
 results in a run-time type error (Section \ref{sec:csem})
 \item we show how to fix this unsoundness issue (Section \ref{sec:csem})
 \item we systematically derive an abstract semantics which represents a correct analyser (we have implemented it in OCaml \cite{ocamlsite})
  (Sections \ref{sec:asem} and \ref{sec:analyzer})
\end{itemize}

In the next sections we first will sketch the type-and-effect system proposed for \links{} (Section ~\ref{sec:rev}), 
then we describe the ideas and the methodology underlying our reconstruction.


\section{Secure Compilation of \links{}}


\label{sec:rev}
Standard web applications have a multi-tier architecture: user interface, application logic and data ma\-nagement are implemented over three different 
tiers. Each tier runs on a different computational environment (web browser, web server and database respectively) characterized by its own language
and its data representation. This heterogeneity gives rise to the problem of \emph{impedance mismatch} \cite{Meijer03programmingwith}: 
because each language has its own data type, data exchanged between tiers of same application have a different representation. 
This problem complicates the development of web  applications because programmers need to define routines to interchange and convert data.
To solve this problem a new class of web languages (multi-tiers languages) have has been developed. These languages allow programmers to blend
server, client and database source code and provide automatic mechanisms for the partition of the application over tiers. 

\links{} is a functional programming language for web applications that belongs to the class of multi-tiers languages.
\links{} enables developers to mix client, server and database source code by delegating the charge of code and data partitioning to the compiler:
from a single source code the compiler generates code for the database back-end, for the server and for the client front-end.

In this way \links{} overcomes the problem of impedance mismatch by abstracting details of a single tier and by supporting an unified programming
model similar to the one used for GUI applications. To realize this cross-tier programming model \links{} exploits the mechanism of the web
continuation \cite{Queinnec:2000:IBE:357766.351243}. These continuations are implemented as closures (expression to be executed plus values of free variables) and are stored
in HTML pages either as hidden fields of forms or as URL parameters. This approach gives rise to security risks since a malicious client may modify
those closures to enforce unexpected computations on the server. 

In particular,  Baltopoulos and Gordon in \cite{baltopoulos09}
have demonstrated that the approach adopted by \links{} of storing unencrypted data on the client is not secure because
an attacker may violate the data secrecy, the data integrity and the control-flow integrity of the application.
To overtake these problems they have proposed a secure implementation of \links{} that includes a compilation strategy based on authenticated encryption
to protect the closures held in the browser and a types-and-effects system to enable source level reasoning about security of web applications. 
This secure implementation has been formalized for \tinylinks{}, a simple subset of \links{}. 

\tinylinks{} is a $\lambda$-calculus augmented with XML values for representing web pages and annotation expressions for expressing safety properties. 
Its syntax is shown in \figurename~\ref{tinylinks:syntax}. 
HTML pages are values created by applying the data constructors $\syntax{Text}$ and $\syntax{Elem}$: 
the first one represents simple text in HTML document, the second one a generic tag element. 
To express links and forms exists two ad-hoc data constructors that contains suspended expressions 
\footnote{we can look at these values as special kinds of functional abstractions.}.
$\syntax{href(E)}$ is a link that, when clicked, evaluates the expression $\syntax{E}$. $\syntax{form(\slist{\indlist{l}{1}{n}},\,E)}$
is a HTML form with a suspended computation (the expression $\syntax{E}$) which requires user input. The input is represented by labels
$\syntax{\slist{\indlist{l}{1}{n}}}$ that will contain the values inserted in the input fields of the form. The evaluation of $\syntax{href}$ and
$\syntax{form}$ can be accomplished by using the operators $\syntax{get}$ and $\syntax{post}$, respectively 
\footnote{we can look at these operations as special kinds of function application.}.
The annotations $\syntax{event\,L}$ and $\syntax{assert\, L}$ have no computational meaning. They allow us to annotate
\tinylinks{} programs with event-based assertions expressing sui\-table safety properties. An expression is safe if whenever an assertion 
$\syntax{assert}\, \syntax{L}$ occurs in the execution, there exists a previous occurrence  
of an event $\syntax{event}\,\syntax{L}$. 
\begin{figure}
  \begin{tabular*}{\textwidth}{ll}
$\syntax{f},\syntax{y},\syntax{x}$
&
Variables
\\
$\syntax{p}$
&
Predicates
\\
$\syntax{c ::=}\, \syntax{Unit} \mid  \syntax{Zero} \mid \syntax{Succ} \mid \syntax{String}$
&
Data constructors
\\
\quad $\mid \syntax{Nil} \mid \syntax{Cons} \mid \syntax{Tuple} \mid \syntax{Elem} \mid \syntax{Text}$
&
\\
$\syntax{g ::=}\, \syntax{+} \mid \syntax{-} \mid \syntax{*} \mid \syntax{/}$
&
Primitive operators
\\
$\syntax{L ::= p\tupleInd{V}{1}{n}}$
&
Events: a predicate and a list of values
\\
$\syntax{V,U ::=}\, \syntax{x} \mid \syntax{c\tupleInd{V}{1}{n}} \mid \syntax{href(E)}$
&
Values
\\
\quad $\mid \syntax{\lambda x_1.\,\ldots,x_n.\, E} \mid \syntax{form(\slist{\indlist{l}{1}{n}},E)}$
&
\\
$\syntax{E ::=}\, \syntax{V} \mid \syntax{var\,x=E_1;E_2} \mid \syntax{g\,(E_1,\,E_2)}$
&
Expressions
\\
\quad $\mid \syntax{V\tupleInd{U}{1}{n}} \mid \syntax{post(\slist{l_1=V_1,\,\ldots,\,l_n=V_n},U)}$
&
\\
\quad $\mid \syntax{get(V)} \mid \syntax{event\,L} \mid \syntax{assert\,L}$
&
\\
\quad $\mid \begin{array}{l}
        \syntax{switch(V) \{} \\
        \quad \syntax{case\, c\tupleInd{x}{1}{n} -> E_1} \\
        \quad \syntax{\_ -> E_2} \\
        \syntax{\}}
       \end{array}$
&
\\
\end{tabular*}
  \caption{Syntax of \tinylinks{}}
  \label{tinylinks:syntax}
\end{figure}

Baltopoulus and Gordon have defined a dependent types-and-effects system to verify that each expression of a program is safe. 
This system is specified by a set of inductively defined typing judgments. These judgments are of the form 
$\Gamma;\syntax{F} |- \syntax{E} \synexp \langle\_:\syntax{T}\rangle\Set{\syntax{F'}}$, where $\syntax{\Gamma}$
is the typing environment, $\syntax{F}$ is the set of events which have occurred and are needed to safe evaluation of the expression $\syntax{E}$ (precondition);
$\syntax{T}$ and $\syntax{F'}$ are, respectively, the type of value and the set of events (post-condition) yielded by the execution of $\syntax{E}$.

The typing rules for the operations $\syntax{get}$ and $\syntax{post}$, for the annotations $\syntax{event}$ and $\syntax{assert}$ and for the function
application are shown in \figurename~\ref{tinylinks:type:rule}. Rule (T-Get) establishes that  
the type assigned to $\syntax{get}$ is $\syntax{xml}$ (that represent the type of a generic HTML tag) with empty effect, 
provided that $\syntax{V}$ is another HTML tag. By (T-Post), the type of $\syntax{post}$ expression is $\syntax{xml}$ with empty effect,
provided that the values associated with submission labels are strings and that $\syntax{U}$ is a HTML tag. 
By (T-Event) $\syntax{event\,L}$ has type $\syntax{unit}$ and effect $\syntax{L}$, 
provided that the values in the event $\syntax{L}$ have a type. Rule (T-Assert) is similar to (T-Event) except that requires 
$\syntax{L \in F}$, that is the precondition of the judgment includes $\syntax{L}$. 
Rule (T-App) is typical for application and shows how the mechanism of the annotations works: the expression is type checked if only if the events
in the precondition $\syntax{F_1}$ of the function have occurred in $\syntax{F}$ with same values. The events generated after application include
the ones of the post-condition of $\syntax{U}$.

\begin{figure}
\[
  \inference[(T-Get)]
    {\Gamma;\syntax{F} |- \syntax{V} \checkval \syntax{xml}}
    {\Gamma;\syntax{F} |- \syntax{get(V)} \synexp \langle\_:\syntax{xml}\rangle\Set{}}
\]
\[
  \inference[(T-Post)]
    {\Gamma;\syntax{F} |- \syntax{V_i} \checkval \syntax{string} & 
     \forallIndex{i}{1}{n} &
     \Gamma;\syntax{F} |- \syntax{U} \checkval \syntax{xml}
    }
    {\Gamma;\syntax{F} |- \syntax{post(\slist{\tuple{l_1=V_1}{l_n=V_n}}\,,U)} \synexp \langle\_:\syntax{xml}\rangle\Set{}}
\]
\[
  \inference[(T-Event)]
    {\Gamma |- \diamond & fv(\syntax{F},\syntax{L}) \subseteq dom(\Gamma) &  \\
     \syntax{L} = \syntax{p\tupleInd{V}{1}{n}} & \Gamma;\syntax{F} |- \syntax{V_i} \synval \syntax{T_i} &
      \forallIndex{i}{1}{n}
    }
    {\Gamma;\syntax{F} |- \syntax{event\,L} \synexp \langle\_:\syntax{unit}\rangle\Set{\syntax{L}}}
\]
\[
  \inference[(T-Assert)]
    {\Gamma |- \diamond & fv(\syntax{F},\syntax{L}) \subseteq dom(\Gamma) & \syntax{L} \in \syntax{F} \\
     \syntax{L} = \syntax{p\tupleInd{V}{1}{n}} & \Gamma;\syntax{F} |- \syntax{V_i} \synval \syntax{T_i} &
      \forallIndex{i}{1}{n}
    }
    {\Gamma;\syntax{F} |- \syntax{assert\,L} \synexp \langle\_:\syntax{unit}\rangle\Set{\syntax{L}}}
\]
\[
  \inference[(T-App)]
    {\Gamma;\syntax{F} |- \syntax{U} \synval \syntax{T} &
     \syntax{T} = \langle\syntax{\tindlist{x}{T}{1}{n}}\rangle\Set{\syntax{F_1}} -> \syntax{T_2\Set{F_2}} &
     fv(\syntax{T}) = \emptyset \\
     \Gamma;\syntax{F} |- \syntax{V_i} \checkval \syntax{T_i} &
     \forallIndex{i}{1}{n} &
      \syntax{\bind{\bind{F_1}{x_1}{V_1}\,\ldots\,}{x_n}{V_n}} \subseteq \syntax{F}}
    {\Gamma;\syntax{F} |- \syntax{U\tupleInd{V}{1}{n}} \synexp \syntax{T_2 \Set{\bind{\bind{\syntax{F_2}}{x_1}{V_1}\,\ldots\,}{x_n}{V_n}}}}
\]
  \caption{Some examples of rules specifying the type-and-effect system for the correspondences analysis.}
  \label{tinylinks:type:rule}
\end{figure}

We say that a web application $\syntax{E}$ is safe if and only if there is a derivation within the types-and-effects system of the judgment 
$\emptyset;\emptyset |- \syntax{E} \checkexp \langle\_:\syntax{xml}\rangle\{\}$, meaning that $\syntax{E}$ is a closed expression which requires
no precondition and which yields a web page without generating further events.

After the definition of typing rules, the standard methodology requires to state and prove the soundness theorem which guarantees the validity of the
analysis with respect to the semantics of the language. Baltopoulus and Gordon adopt a different approach by translating each \tinylinks{} expression
to an expression of a concurrent $\lambda$-calculus with \emph{refinement types}. This translation hides the details and the properties of the defined 
types-and-effects system, in particular the soundness. For instance, the expression $\syntax{get(Text("Hello!"))}$ is safe because a derivation 
exists for the judgment $\emptyset;\emptyset |- \syntax{get(Text("Hello!"))} \checkexp \langle\_:\syntax{xml}\rangle\Set{}$. 
However, we will show in the next section 
that the proposed types-and-effects system is not sound because, even if this expression is type checked, its evaluation results in a run-time type error.


 \section{A Denotational Semantics for \tinylinks{}}

\label{sec:csem}
In this paper we adopt the approach described by Cousot in \cite{Cousot97-1}.
We define a denotational semantics for \tinylinks{}, by considering it as an untyped
$\lambda$-calculus. Furthermore, since we deal with effects, we explicitly consider assertions of events.
To this purpose we introduce a special environment (\emph{events environment}) which will store occurred events. The semantics
of $\syntax{assert\,q\tupleInd{V}{1}{n}}$ will require checking that $\syntax{q}$ is bound in this environment 
to values $\syntax{\indlist{V}{1}{n}}$. If this check succeeds, the evaluation yields a $\syntax{Unit}$ value, otherwise
a ``sentinel`` value indicating an error.

For the sake of simplicity, we restrict the values in an event to integers only. 
We will also assume that functions have a single argument and predicates in events are bound to a single value.
Since we regard \tinylinks{} an untyped $\lambda$-calculus, we define the semantics domain of values ($Eval$) as a recursive sum of cpos, by
using the inverse limit construction described in \cite{schmidt86}. Each element of this sum represents a specific class of values. For instance, $\Int{}$
is the set of integers; $U$ and $S$ are singletons of the $\syntax{unit}$ value and the error value
\footnote{this value is used to show a run-time type error};
$EEnv -> Eval -> (Eval \times EEnv)$, $EEnv -> (Eval \times EEnv)$ and $EEnv -> \slist{Eval} -> (Eval \times EEnv)$ are the sets of
the denotations of functions, links and forms, respectively.

The environment ($Env$) is a function from identifier ($Ide$) to values ($Eval$). The events environment ($EEnv$) maps predicates ($Pred$)
to pairs formed by an element of $Dval$ and an element of $Mark$. $Dval$ denotes values which can occur in an event
\footnote{in the following we will call them denotable values}. 
$Mark$ is the state of an event: $E$ indicates that the event has occurred, $EA$ that has occurred and has been asserted, 
$A$ that has only been asserted.

We define two semantic functions $\vsem{-}{}{}\colon \syntax{VAL} -> Env -> EEnv -> Eval$ for values and
$\ssem{-}{}{}\colon \syntax{EXP} -> Env -> EEnv -> (Eval \times EEnv)$ for expressions. 
The semantics of values is straightforward, because we only need to construct the corresponding denotation. Some examples of semantic equation are
shown in \figurename~\ref{csem:val}. In the definition, we use injections into $Eval$ (like $Unit$, $Href$, $Fun$),
continuous semantic operators (like $bindList$) and a meta-language which includes:
\begin{itemize}
 \item $if\,e_1\,then\,e_2\,else\,e_3$ (conditional);
 \item $let\,x=\,e_1\,in\,e_2$ as a cleaner notation for $((\lambda x.e_2)\, e_1)$;
 \item $let^\star\,x=\,e_1\,in\,e_2$ for $((\lambda x.e_2)^\star\, e_1)$;
 \item $case\,e_1\,of\,in_1(x_1) \to e_2 \,\,\_ \to e_3$ for $[\lambda x_1.\,e_1,\,\lambda x_2.e_3,\,\ldots,\, \lambda x_2.e_3]$;
 \item $let\,(x_1\,x_2)\,=e_1\,in\,e_2$ for $let\,y=e_1\,in\,let\, x_1\,=\pi_1(y)\,let\,x_2\,=\pi_2(y)\,in\,e_2$;
\end{itemize}
where $\pi_i$, $\lift{-}$, $\star$ and $[-,\ldots,-]$ are the standard operators for product, lifting and sum of cpos \cite{Winskel99}.
\begin{figure}[!ht]
\begin{align*}
\vsem{\lambda x.\,E}{\rho}{\phi}\,\,=\,\,
& \lift{Fun(\lambda \phi'.\,\lambda v.\,\ssem{E}{\bind{\rho}{x}{v}}{\phi'})}\\
\vsem{Href(E)}{\rho}{\phi}\,\,=\,\,
&\lift{Href(\lambda \phi'.\,\ssem{E}{\rho}{\phi'})}\\
\vsem{Form(ll,\,E)}{\rho}{\phi}\,\,=\,\,
&\lift{Form(\lambda \phi'.\,\lambda vl.\,let^\star\,\,\rho' = bindList(\rho,\,ll,\,vl)\,\,in\,\,\ssem{E}{\rho'}{\phi'})}
\end{align*}  
  \caption{Examples of semantic equations for values.}
  \label{csem:val}
\end{figure}

The semantics of expressions is similar to the one of the untyped 
$\lambda$-calculus. The most interesting cases of semantic equations are shown in \figurename~\ref{csem:exp} and 
below we give some comments about them.

The semantics of $\syntax{get(V)}$ asks to evaluate $\syntax{V}$; if the evaluation results into the denotation of a link
($Href(f)$), we evaluate the corresponding suspended expression (the closure $f$), otherwise we return an error value.

The semantics of $\syntax{post(VL,\,V)}$ is similar: if the evaluation of $\syntax{V}$ is a form ($Form(f)$) and the evaluation of $\syntax{VL}$
is a list of strings, we return the result of the application of the functional value $f$ to the denotation of
$\syntax{VL}$ and to the current events environment $\syntax{\phi}$.

The semantics of $\syntax{event\,q(V)}$ requires the evaluation of $\syntax{V}$; if the produced value is an integer, 
we create a new binding for the predicate $\syntax{q}$ in $\syntax{\phi}$ and return a unit value otherwise we raise an error.

The semantics of $\syntax{assert\,q(V)}$ is similar, but requires the evaluation of $\syntax{V}$ to be equal to the value bound 
to the predicate $\syntax{q}$
in $\syntax{\phi}$. In this case we update the state of the event in $\syntax{\phi}$ and return a unit value.
\begin{figure}[!ht]
\begin{align*}
\ssem{get(V)}{\rho}{\phi}\,\,=\,\,
&let^\star\,\,v'\,= \vsem{V}{\rho}{\phi}\,\,in\\
&case\,\,v'\,\,of\\
&\quad Href(f) -> f\,\phi\\
&\quad \_ -> (\lift{WrongValue()},\,\iota)\\
\ssem{post(VL,\,V)}{\rho}{\phi}\,\,=\,\,
&let^\star\,\,v'\,=\vsem{V}{\rho}{\phi}\,\,in\\
&let^\star\,\,v_2\,=checkStringList(map\,(\lambda x.\,\vsem{x}{\rho}{\phi})\,\syntax{VL})\,\,in\\
&case\,\,v'\,\,of\\
&\quad Form(f) ->\,case\,\,v_2\,\,of \\
&\qquad\qquad\qquad\qquad V(vl) -> f\,vl\,\phi\\
&\qquad\qquad\qquad\qquad \_ -> (\lift{WrongValue()},\,\iota)\\
&\quad \_ -> (\lift{WrongValue()},\,\iota)\\
\ssem{event\,q(V)}{\rho}{\phi}\,\,=\,\,
&let^\star\,\,d\,= evalToDval(\vsem{V}{\rho}{\phi})\,\,in\\
&if\,\,d = dint(n)\,\,then\\
&\quad (\lift{Unit()},\,\bind{\phi}{\syntax{q}}{(d,\,E)})\\
&else\\
&\quad (\lift{WrongValue()},\,\iota)\\
\ssem{assert\,q(V)}{\rho}{\phi}\,\,=\,\,
&let^\star\,\,ev\,=evalToDval(\vsem{V}{\rho}{\phi})\,\,in\\
&let\,\,(ev',\,m)\,=\phi\,\syntax{q}\\
&if\,\,ev\,=\,ev'\,\,then\\
&\quad (\lift{Unit()},\,\bind{\phi}{\syntax{q}}{(ev',\,EA)})\\
&else\\
&\quad (\lift{WrongValue()},\,\iota)
\end{align*}
  \caption{Examples of semantic equations for expressions.}
  \label{csem:exp}
\end{figure}

By using the semantic equation of $\syntax{get}$ we prove that the evaluation of the expression $\mathrel{\syntax{get(}}
\mathrel{\syntax{Text(}}\syntax{"Hello!"}\mathrel{\syntax{)}}\mathrel{\syntax{)}}$
results in a run-time type error (the value $\lift{WrongValue()}$) because the denotation of $Text("Hello!")$ is not a link.
Although a link is an XML value, it is different from other XML va\-lues because it is a special kind of functional
abstraction. Notice that the type-and-effect system proposed for \tinylinks{} does not handle this special nature of links correctly, 
because it assigns the same type to the all XML values. Note that the same remark can be made for forms. 
Our above arguments demonstrate that the types-and-effects system of \cite{baltopoulos09} is unsound because exists
an expression which is type checked but its evaluation yields yet a run-time type error. 
We argue that the solution to this problem is to use a type system with subtypes. 
For the sake of simplicity, in our reconstruction we will not use subtypes, but we will instead define two ad-hoc 
types for forms and links which will handled so as have a sound analysis.


\section{An Abstract Semantics for Inference of Types and Effects}


\label{sec:asem}

Following the classical methodology of abstract interpretation, once we have defined a concrete semantics, 
we need to define a collecting semantics by extending $\vsem{-}{}{}$ and $\ssem{-}{}{}$ to the powerset.

The concrete semantics properties, which we are interested in, are the types and the event-based annotations. We need to define a suitable
domain for both. One possibility is to define the abstract domain as the set of Hindley's monotypes (terms) with variables  
\cite{Hindley69,DamasMilner82,Cousot97-1,Monsuez92,Gori02}. However, this is not possible, since types
are annotated by effects.  For example, a function type will have the form $\syntax{T_1\Set{F_1} -> T_2\Set{F_2}}$,
where $\syntax{F_1}$ are the events which have to be occurred before the function application, whereas $\syntax{F_2}$ are 
the events which we can consider occurred afterwards. Hence, we need to define a domain of annotated types. 
The main problem is that the algebra of annotated terms is not free. In fact,
two types can be identified even if their syntax is different. For example, the types 
$\syntax{xml}\syntax{\Set{q(10),\,p(1)} -> \syntax{xml}\syntax{\Set{}}}$ and $\syntax{xml}\syntax{\Set{p(1),\,q(10)} 
-> \syntax{xml}\syntax{\Set{}}}$ 
have a different
representation, but they are equal because the effects $\Set{q(10),\,p(1)}$ and $\Set{p(1),\,q(10)}$ denote the same set. 
Therefore, we cannot use a syntactic unification algorithm \cite{Lassez88} to solve equations between terms. 

One solution would be to use an algorithm for unifying terms in non-free algebras (semantic unification). 
Such algorithms do exist \cite{baader94}, but they are not usable in practice.

Our reconstruction does not rely on semantic unification but on another approach described in \cite{NiNi05ppa}.
This approach exploits special annotated types (simple types), where annotations are replaced by variables (annotation variables), 
whose values have to satisfy some constraint.  For example, the annotated type $\syntax{xml\Set{q(10),\,p(1)} -> xml\Set{}}$
becomes $\syntax{xml(\alpha) -> xml(\beta)}$, where $\alpha$ and $\beta$ are the minimal annotations $A$ and $B$ which satisfy 
the constraints $A \supseteq \Set{q(10),\,p(1)}$ and $B \supseteq \Set{}$, respectively. 
The algebra of simple types is free. Hence, the introduction of a new kind of variable in terms 
requires a simple variation of the unification algorithm: an annotation variable unifies with another annotation variable only. 

However, this solution is not completely adequate to define an abstract domain for the properties which we are concerned with, because the
effects depend on the values. Hence we need to include them in the abstract domain.
Since events in the precondition and post-condition of a function type may depend on the value bound to a formal parameter we need to remember it.
We then introduce in the set of terms another kind of variables, called identifier variables. 
Identifier variables are handled by simple modification of the unification algorithm: 
an identifier variable unifies with another identifier variable only.

The domain of abstract values will contain also substitutions as in \cite{Gori02}. %
The role of substitutions can be explained as follows. At some point in the evaluation of the abstract semantics 
(for example, in the semantics of function abstraction), we will introduce new type variables, with the meaning "any possible type". 
During the evaluation (for example, of the function body), this information 
will be subject to instantiations, computed by unifications and represented as an idempotent substitution. 
Since the abstract semantic evaluation functions are defined by structural recursion, the easiest way to provide 
the instantiation information to the caller is to include it in the returned value.

Although we have now all necessary information for defining an adequate abstract domain, there is a problem concerning the representation of 
effects in the constraints. Intuitively we can simply represent them by using a set of pairs, where the first component is the predicate and 
the second one is the denotable value. The problem is in partial order, since we should consider both set inclusion and the relative 
precision of denotable values. We can achieve this by using power domains \cite{Gunter90,schmidt86}. 
We use a different approach: we define an effect as a function from predicates to denotable values (we will name it correspondence function). 
We can then represent constraints by splitting them in two parts: the first part is
a set of pairs (annotation variable, predicate) and the second one is a correspondence function.

Let $V_t$ be a countable set of type variables, $V_a$ be a countable set of annotation variables, $Ide$ be a countable set of identifier variable
($V_t \cap V_a \cap Ide = \emptyset$) and $\Sigma = \{unit:0, int:0, string:0, xml:1, link:1, form:1, list:1, fun:5\} 
\cup \{tuple_n : n \mid  n \geq 2\}$ be a numerable set of function symbol, $T_s$ is the set of terms with variables $V_t \cup V_a \cup Ide$ modulo
renaming, ordered by the inverse instance relation. 
It is worth noting that we have introduced two new types $form$ and $link$ in order to solve the problem relating forms and links which we described 
in Section \ref{sec:csem}. Furthermore we will use annotation variables in $xml$, $link$, $form$ and $fun$ only; 
in $fun$ there are two annotation variables representing the precondition and the post-condition respectively. We further assume that the first
argument of $fun$ is an identifier variable. We obtain $TypeS$ by lifting $T_s$ with idempotent substitutions \cite{Gori02} and by adding a new bottom 
element $Notype$.

As we described above, the first part of a constraint is a pair (annotation variable, predicate): $(\delta,\,q)$ means that 
the predicate $q$ is in the effect represented by the variable $\delta$. We use inverse inclusion as partial order: 
if $C_1$ is included in  $C_2$, then $C_1$ has less information than $C_2$, hence, its value is less precise.
Let $V_a$ be the set of annotation variables and $Pred$ be the set of predicates. We define $Constr = \power{V_a \times Pred}$.
The second part of a constraint is a correspondence function whose domain is $TPred = Pred -> Dval$ ordered by using the dual of usual partial order. 
We assume that $cb \colon \power{TPred} -> TPred$ is the glb operator and $\zeta$ is the bottom element.

The domain of abstract values is $TypeA = TypeS \times Dval \times Constr \times TPred$. In the following, we will denote by $Error$ the
bottom element of this domain. 

The domain of abstract environment (type environment) is $AEnv = Ide -> TypeA$. 
We are now in the position to define our abstract domains $AV = AEnv -> EEnv -> TypeA$ for values and 
$AE = AEnv -> EEnv -> (TypeA \times EEnv)$ for expressions.

To relate the abstract domain to the concrete one we need to define a Galois connection. In \cite{galletta10} we formally built this
connection in in various steps, by using properly defined representation functions \cite{NiNi05ppa} and propositions.

Some examples of abstract semantic equations are shown in Figures \ref{fig:asem1}, \ref{fig:asem2} and \ref{fig:asem3}.
\begin{figure}[!ht]
\begin{align*}
& \avsem{href(E)}{\rho}{\phi} \,=\,\\
& \qquad \gamma \in V_a \quad fresh\\
& \qquad let\, ((ts,\,\_,\,C,\,f),\,\phi') \,=\, \assem{E}{\rho}{\phi}\,in\\
& \qquad let\, A\, = assert(\phi',\,\phi) \,in\\
& \qquad let\, E\, = event(\phi',\phi) \,in\\
& \qquad if\,\,E = \emptyset \land ts \neq NoType\,\,then\\
& \qquad \quad case\, mgu(\Set{ts.t= xml(\gamma)} \cup ts.\theta)\,of\\
& \qquad \quad \quad S(\theta) -> let\,\,C'\,=\, C \cup \Set{(\gamma,\,q) \mid q \in A}\\
& \qquad \quad \quad \quad let\,\,f'\, =\, cb\Set{f,\,\eenvTotpred(diff(\phi',\,\phi))}\,in\\
& \qquad \quad \quad \quad ((\theta(link(\gamma)),\,\theta),\,nodval,\,\theta(C'),\,\theta(f'))\\
& \qquad \quad \quad \_ -> Error\\
& \qquad else \\
& \qquad \quad Error\\
& \avsem{\lambda x.\,E}{\rho}{\phi} \,=\, \\
& \qquad \alpha \in V_t \quad \gamma_1,\,\gamma_2 \in V_a \quad fresh \quad \epsilon \quad identity\,\,\, substituition\\
& \qquad let\,\,((ts,\,\_,\,C_1,\,f_1),\,\phi')\,=\, \assem{E}{\bind{\rho}{x}{((\alpha,\,\epsilon),\,var(x),\,\emptyset,\,\zeta)}}{\phi}\,in\\
& \qquad if\,\,ts \neq NoType\,\,then\\
& \qquad \quad let\,\,\phi_d = \theta(\phi)\\
& \qquad \quad let\,\, C' = \Set{(\gamma_1,\,q) \mid q \in assert(\phi',\,\phi_d)}\,in\\
& \qquad \quad let\,\, C'' = \Set{(\gamma_2,\,q) \mid q \in event(\phi',\,\phi_d)}\,in\\
& \qquad \quad let\,\, f_2 = \eenvTotpred(diff(\phi',\,\phi_d))\,in\\
& \qquad \quad \quad ((\theta(fun(x,\,\alpha,\,\gamma_1,\,ts.t,\,\gamma_2)),\,\theta),\\
& \qquad \quad \quad \quad nodval,\,\theta(C_1 \cup C' \cup C''
),\,\theta(cb\Set{f_1,\,f_2}))\\
& \qquad else\\
& \qquad \quad Error
\end{align*}
\caption{The abstract semantics of links and functional abstractions.}
\label{fig:asem1}
\end{figure}
In these definitions, we assume to have a function $mgu$, which, given a set of term equations, computes a solution by using the unification algorithm. 
If there exists a solution, it returns the unifier $S(\theta)$; otherwise, it returns $F$ to denote failure. 
The set of equations is denoted by $\{t_1 = t'_1,\,\ldots,\,t_n = t'_n\}$.
Since idempotent substitutions are isomorphic to solved form equations, we will use $\{t_1 = t'_1,\,\ldots,\,t_n = t'_n\} \cup \theta$
to refer the union of equations in  $\{t_1 = t'_1,\,\ldots,\,t_n = t'_n\}$ and equations defined by $\theta$.
For the sake of simplicity, the components of the elements of the domain $TypeS$, will be identified by a notation 
similar to the one used to access the fields of a structure in an imperative language. 
Given $ts=(t',\,\theta') \in TypeS$, then $ts.t = t'$ and $ts.\theta = \theta'$.

Given an element $C$ of $Constr$ and a substitution $\theta$, we will denote by $\theta(C) = \{(\theta(\delta),\,l) \mid (\delta,\,l) \in C\}$
the pair obtained by applying $\theta$ to all the annotation variables in $C$.

Given a correspondence function $f \in TPred$ and a substitution $\theta$, we define $\theta(f) = \lambda q. \theta(f\,q)$,
where if $d \neq var(x)$ for some $x$ then $\theta(d) = d$.

\begin{figure}[!ht]
\begin{align*}
& \assem{get(V)}{\rho}{\phi} \,=\,\\
& \qquad \gamma \in V_a \quad fresh\\
& \qquad let\,\,(ts,\,d,\,C,\,f)\,=\avsem{V}{\rho}{\phi}\,in\\
& \qquad if\,ts \neq NoType\,then\\
& \qquad \quad case\,\,mgu(\Set{ts.t = link(\gamma)} \cup ts.\theta)\,\,of\\
& \qquad \quad \quad S(\theta) -> let\,\, C' \,=\, \Set{(\theta(\gamma),\,q) \in \theta(C)}\,in\\
& \qquad \quad \quad \quad if\, check(\theta(f <- C'),\,\phi)\,\,then\\
& \qquad \quad \quad \quad \quad (((\theta(xml(\gamma)),\,\theta),\\
& \qquad \quad \quad \quad \quad \quad \quad \quad nodval,\,\theta(C) \setminus C',\,\theta(f \downarrow C)),\phi)\\
& \qquad \quad \quad \quad else\\
& \qquad \quad \quad \quad \quad (Error,\,\iota)\\
& \qquad \quad \quad \_ -> (Error,\,\iota)\\
& \qquad else\\
& \qquad \quad (Error,\,\iota)\\
& \assem{E_1\,E_2}{\rho}{\phi} \,=\,\\
& \qquad x \in Ide\,\,\alpha_1 \in V_t\,\,\gamma_1,\,\gamma_2 \in V_a \quad fresh\\
& \qquad let\,\,((ts_1,\,\_,\,C_1,\,f_1),\,\phi_1)\,= \assem{E_1}{\rho}{\phi}\\
& \qquad let\,\,((ts_2,\,d_2,\,C_2,\,f_2),\,\phi_2)\,= \assem{E_2}{\rho}{\phi_1}\\
& \qquad if\,\,ts_1 \neq NoType \land ts_2 \neq NoType\,\,then\\
& \qquad \quad case\,\,mgu(\Set{ts_1.t = fun(x,\,\alpha,\,\gamma_1,\,ts_2.t,\,\gamma_2)} \cup \\
& \qquad \qquad \qquad \qquad \cup ts_1.\theta \cup ts_2.\theta)\,\,of\\
& \qquad \quad \quad S(\theta) -> let\,\, C'\,=\,\Set{(\delta,\,q) \in \theta(C_1) \mid \delta \in prvar(\theta(ts_1.t))}\,in\\
& \qquad \qquad \qquad \quad\,\,  let\,\, C''\,=\,\Set{(\delta,\,q) \in \theta(C_1) \mid \delta \in psvar(\theta(ts_1.t))}\,in\\
& \qquad \qquad \qquad \quad\,\,  let\,\, f'_1\,=\,\theta(f_1)[\theta(x),\,d_2]\,in\\
& \qquad \qquad \qquad \quad\,\,if\,\,check(\theta(f'_1 <- C'),\,\phi_2)\,\,then\\
& \qquad \qquad \qquad \qquad (((\theta(ts_2.t),\,\theta),\,\top,\,\theta(C_1 \cup C_2)\setminus (C' \cup C'),\,\\
& \qquad \qquad \qquad \qquad \qquad cb\Set{\theta(f_1) \downarrow (C' \cup C''),\,\theta(f_2)}),\, incl(\phi_2,\,(\theta(f'_1) <- C'')))\\
& \qquad \qquad \qquad \quad\,\,else\\
& \qquad \qquad \qquad \qquad (Error,\, \iota)\\
& \qquad \quad \quad \_ -> (Error,\, \iota)\\
& \qquad else\\
& \qquad \quad (Error,\, \iota)
\end{align*}
\caption{The abstract semantics of \code{get} expression and function application.}
\label{fig:asem2}
\end{figure}
Furthermore we assume that for $f \in TPred$ and $C \in Constr$ $f \downarrow C$ and $f <- C$ are the correspondence functions achieved by removing from $f$
the predicates occurring and not occurring in $C$ respectively; that $f[x,\,d]$ for $\syntax{x \in Ide}$, $d \in DVal$ and $f \in TPred$ 
is the correspondence function achieved by binding $d$ to all predicates which are bound to $var(x)$ in $f$;
that given a $f \in TPred$ and $\phi \in EEnv$ the function $check(f,\,\phi)$ 
returns $true$ if the events represented by $f$ have been occurred in $\phi$, $false$ otherwise;
that for $\phi \in EEnv$ $eenvToTPred(\phi)$ is the correspondence function achieved from $\phi$; that given 
$\phi_1$,$\phi_2$ $assert(\phi_2,\,\phi_1)$ is the set of predicates of events asserted in $\phi_2$ but not in $\phi_1$,
that $event(\phi_2,\,\phi_1)$ is the set of predicates of events generated in $\phi_2$ but not in $\phi_1$ and that $diff(\phi_2,\,\phi_1)$
is the events environment which contains the events of $\phi_2$ which are not in $\phi_1$ and the events of $\phi_1$ which changed their
value or state in $\phi_2$. Furthermore we assume that, given $t \in T_s$, $prvar(t)$ and $psvar(t)$ denote the set of annotation variables
of $t$ for preconditions and post-conditions, respectively.
\begin{figure}[!ht]
\begin{align*}
& \assem{assert\,q(V)}{\rho}{\phi} \,=\,\\
& \qquad let\,\,(ts,\,d,\,C,\,f)\,=\,\avsem{V}{\rho}{\phi}\,in\\
& \qquad if\,\,ts \neq NoType \land (d = nint(n) \lor d = var(x))\,\,then\\
& \qquad \quad case\,\,mgu(\Set{ts.t = int} \cup ts.\theta)\,\,of\\
& \qquad \quad \quad S(\theta) -> if\,\,q \notin dom(\phi) \lor \pi_1(\phi(q)) = d \,\,then\\
& \qquad \qquad \qquad \qquad \quad (((unit,\,\theta),\,nodval,\,\theta(C),\,\theta(f)),\,\bind{\phi}{q}{(d,\,A)})\\
& \qquad \qquad \qquad \qquad else\\
& \qquad \qquad \qquad \qquad \quad (Error,\,\iota)\\
& \qquad \quad \quad \_ -> (Error,\,\iota)\\
& \qquad else\\
& \qquad \quad (Error,\,\iota)\\
& \assem{event \,q(V)}{\rho}{\phi} \,=\,\\
& \qquad let\,\,(ts,\,d,\,C,\,f)\,=\,\avsem{V}{\rho}{\phi}\,in\\
& \qquad if\,\,ts \neq NoType \land (d = nint(n) \lor d = var(x))\,\,then\\
& \qquad \quad case\,\,mgu(\Set{ts.t = int} \cup ts.\theta)\,\,of\\
& \qquad \quad \quad S(\theta) -> if\,\, q \notin dom(\phi) \lor \phi(q) = (d,\,T)\,\,then\\
& \qquad \qquad \qquad \qquad \quad (((unit,\,\theta),\,nodval,\,\theta(C),\,\theta(f)),\,\bind{\phi}{q}{(d,\,E)})\\
& \qquad \qquad \qquad \qquad\,\,else\\
& \qquad \qquad \qquad \qquad \quad (Error,\,\iota)\\
& \qquad \quad \quad \_ -> (Error,\,\iota)\\
& \qquad else\\
& \qquad \quad (Error,\,\iota)
\intertext{for some $n \in \Int$, $x \in Ide$ and where $T \in \Set{E,\,EA}$}
\end{align*}
\caption{The abstract semantics of \code{event} and \code{assert} annotations.}
\label{fig:asem3} 
\end{figure}

The semantics of links consists in the evaluation of the expression $\syntax{E}$. If in this evaluation no errors ($ts \neq NoType$) and
no new events (this is required by the rule described in Section \ref{sec:rev}) occur, then we check that the computed value has type $xml$.
Since in our reconstruction $xml$, $link$ and $form$ are different types without any relation, this check rejects all the expressions which
return a value of type $link$ or $form$. Although this behavior may seem too restrictive, because it rejects some legal expressions 
like $\syntax{href(href(Text("Hello")))}$, it guarantees us safety and simplicity in the management of these diffe\-rent and unrelated types.
If this check has success, we return an abstract value where the simple type is $link$ and the 
constraint is risen by properly extending the result of the evaluation of $\syntax{E}$.

The semantics of forms is similar. We evaluate $\syntax{E}$ in a type environment where the labels $\syntax{ll}$ are bound to the
abstract value with simple type $string$ and constraint empty and we return an abstract value where the simple type is $form$.

The semantics of functional abstraction consists in the evaluation of the body $\syntax{E}$ in a type environment, where the formal parameter
$\syntax{x}$ is bound to a generic type. If in this evaluation no errors occur, we compute the events which 
are included in the precondition (represented by $C'$ and $f_2$) and in the post-condition (represented by $C''$ and $f_2$). We return an abstract 
value where the simple type is obtained by applying the substitution $ts.\theta$ to the functional type $(fun(x,\,\alpha,\,\gamma_1,\,ts.t,\,\gamma_2))$
and the constraint is obtained by combining $C$ with $C'$ and $C''$ and $f$ with $f_2$.

The semantics of $\syntax{get}$ requires the evaluation of $\syntax{V}$ to be successful and yields a value of type $link$. If the preconditions
are satisfied, that is if they are in $\phi$ and have occurred before, we construct an abstract value where the simple
type is $xml$ and the constraint is obtained from the one returned by the evaluation of $\syntax{V}$ by removing 
the information about preconditions. The pair which is returned has in the first component this abstract value and in the second one the events
environment $\phi$. This is correct because the semantics of $\syntax{href}$ guarantees that no new events have occurred during the evaluation of the
suspended expression.

The semantics of $\syntax{post}$ is similar except that we ask that the elements of list $\syntax{VL}$ are strings and
that the value yielded by the evaluation of $\syntax{V}$ has type $form$.

In the semantics of function application we evaluate the sub-expressions $\syntax{E_1}$ and $\syntax{E_2}$: if both evaluations do not
produce errors, we check that the simple type of $\syntax{E_1}$ is a function type where the argument has the simple type of $\syntax{E_2}$ and 
that the precondition of function is satisfied in the events environment $\phi_2$, obtained from evaluating both the sub-expressions. In order to perform
this last check, we substitute the denotable value bound to $\syntax{x}$ in $f_1$ by the one returned by the evaluation 
of $\syntax{E_2}$. Then, by using the function $check$, we ask that the events required by the function body are  in $\phi_2$. If we 
succeed, we construct an abstract value where the simple type is $\theta(\alpha)$ and the constraint is
obtained by composing those returned by the evaluation of the sub-expressions, where the events of preconditions and post-conditions are removed. 
We return a pair composed by this abstract value and by the events environment $\phi_2$ extended with the events of the post-condition of the function.

The semantics of $\syntax{assert}$ consists in the evaluation of $\syntax{V}$. If it yields an abstract value whose simple type is $int$ and whose 
denotable value is a specific integer or a specific identifier, we check that there is in $\phi$ at most the same event which we are generating.
In this way we are sure that it is impossible to change the value bound to a predicate. If this check has success, we build an abstract value where the 
simple type is $unit$ and the constraint is the one returned by the evaluation of $\syntax{V}$. This abstract value
is the first component of returned pair; the second component consists of the events environment $\phi$ extended with the new event.

The semantics of $\syntax{event}$ is similar except that we ask that, if the event is in $\phi$, then its state has to be either $E$ or $EA$.


\section{Implementation and Examples}

\label{sec:analyzer}
Both the concrete and the abstract semantics have been implemented as OCaml \cite{ocamlsite} programs. 
The language provides a feature, the mechanism of functors, which allows us to have a 
unique semantic function (realised by the functor \code{Semantics}),  
parametrized with respect to the primitive operations and the semantic domain. 
We can thus construct the concrete semantics interpreter, which executes programs, and the abstract interpreter, 
which analyzes programs in terms of types and effects, by instantiating the same functor \code{Semantics}.

Programs are represented in abstract syntax, although, for the sake of simplicity, we will use in the following \links{}-like syntax.
For example, the expression
\begin{center}
\begin{varwidth}{\linewidth}
\begin{verbatim}
fun buy(value, dbpass) {
  var _ = assert PriceIs(value);
    Text("Hello")
}
\end{verbatim}
\end{varwidth}
\end{center}
defines a function which requires that the event \code{PriceIs(value)} has occurred and which returns an XML value.
The result of its evaluation by the abstract semantics interpreter is
\begin{center}
\begin{varwidth}{\linewidth}
\begin{verbatim}
(type - :
 Function(_#value#var0_, Integer(), _annvar0_,
   Function(_#dbpass#var1_, _typevar1_, _annvar2_, 
            Xml(_annvar4_), _annvar3_),
   _annvar1_)
 No_dval [(_annvar2_,PriceIs)] {PriceIs -> _#value#var0_}, {})
\end{verbatim}
\end{varwidth}
\end{center}
meaning that the computed type is a function type whose first argument has a type integer and the se\-cond one has type variable
\footnote{since the \code{dbpass} parameter is not used in the body, the analyzer cannot compute a more precise type} 
where the precondition (represented by the annotation variable \code{\_annvar2\_}) 
includes the event composed by the predicate \code{PriceIs} and the value bound to the first formal parameter.
If we give a value (for example 5) to the first parameter, the abstract semantics is
\begin{center}
\begin{varwidth}{\linewidth}
\begin{verbatim}
(type - :
 Function(_#dbpass#var3_, _typevar3_, _annvar7_, 
          Xml(_annvar9_), _annvar8_)
 Unknown [(_annvar7_,PriceIs)] {PriceIs -> 5}, {})
\end{verbatim}
\end{varwidth}
\end{center}
that is the computed type is a specialization of that one computed for \code{buy} where the predicate \code{PriceIs} is bound
to the value \code{5} in the precondition. The abstract semantics of the application of the function \code{buy} to \code{5} and \code{"a"} is
an error
\begin{center}
\begin{varwidth}{\linewidth}
\begin{verbatim}
  Exception: No_type "apply_fun: no preconditions"
\end{verbatim}
\end{varwidth}
\end{center}
because we are applying a function whose precondition is not satisfied.


\section{Conclusions}
We have described how to reconstruct a types-and-effects system, proposed to handle some security issues in \links{}, as 
an abstract interpretation of a denotational semantics which explicitly models the types and the effects. By our reconstruction
we have precisely defined the relation between the semantics and the analysis, we have systematically constructed a correct analyser 
and we have shown that the proposed types-and-effects system was not sound. We have stressed that the unsoundness derived from the fact 
of considering forms and links as simple XML values forgetting their own differentiating features. In our reconstruction we have solved 
this problem by using two new specific types and we have managed them in ad-hoc manner. 
We plan to extend our reconstruction to consider a type system with sub-types so as to be able to manage links and 
forms in a more uniform and elegant way and to use additional values in the effects.
  
One advantage of abstract interpretation approach on the type system approach is that the analysis is directly derived from the 
semantics and is sound by construction. This forces one to tackle from the very beginning subtle problems such as the ones described in Section 
\ref{sec:csem} that might only be revealed while trying to prove the soundness theorem following the type system approach. On the other hand we 
have shown that abstract interpretation can easily handle extensions of types, such as types and effects. 
There is only one example in the literature of an
abstract interpretation reconstruction of a type and effect static analysis \cite{Vouillon95typeand}.

\bibliographystyle{eptcs}

\bibliography{reference.bib}


\end{document}